\documentstyle[preprint,prb,aps]{revtex}
\tightenlines
\begin{document}
\draft
\title{Transverse dipole spin modes in quantum dots}

\author{E. Lipparini$^1$, M. Barranco$^1$\cite{perm}, A. Emperador$^2$, 
M. Pi$^2$, and Ll. Serra$^3$}

\address{$^1$Dipartimento di F\'{\i}sica, Universit\`a di Trento. I-38050
Povo, Italy}

\address{$^2$Departament
d'Estructura i Constituents de la Mat\`eria, Facultat de F\'{\i}sica,
\\Universitat de Barcelona. E-08028 Barcelona, Spain}

\address{$^3$Departament de F\'{\i}sica,
Universitat de les Illes Balears, E-07071 Palma de Mallorca, Spain}
\date{\today}

\maketitle

\begin{abstract}
We have carried out a systematic analysis of the transverse dipole
spin response of a large size quantum dot within time-dependent
current density functional theory. Results for magnetic fields
corresponding
to integer filling factors are reported, as well as a comparison
with the longitudinal dipole spin response.
As in the two dimensional electron gas, the spin response  at
high spin magnetization is dominated by a low energy transverse mode.
\end{abstract}


\narrowtext

\section{Introduction}

Resonant inelastic light scattering has become a very useful tool to
study quantum dot (QD) excitations\cite{Str94,Loc96,Sch96,Sch98}.
It had been extensively employed before to study electron
single-layers and quantum wires in semiconductor heterostructures
(see for example Refs. \onlinecite{Pin89a,Sch94}),
and currently it is also being
applied to study electron multi-layers\cite{Pel97,Pla97}. We refer the
reader to Refs. \onlinecite{Pin89b,Pin91} for a review of this experimental
technique.

A major advantage of resonant inelastic light scattering over far infrared (FIR)
optical absorption techniques used in the past\cite{Sik89,Dem90} is that
the former allows to disentangle and identify, using polarization selection
rules in the backwards geometry\cite{Pin89b}, charge density (CDE), spin
density (SDE) and single-particle (SPE) electron excitations in the same
sample, whereas FIR absorption is only sensitive to charge density
excitations. In inelastic light scattering, when the polarizations of the
incoming and scattered photon are parallel (polarized geometry)
 CDE's are observed, whereas
when the polarizations are perpendicular (depolarized geometry) SDE's
dominate the spectrum. This is due
to the structure of the scattering cross section, which besides the charge or
spin electronic strength function contains the scalar or vector product
of the photon polarizations, respectively\cite{Kat85,Luo93}.
Yet, CDE's are seen with some intensity in the depolarized
spectrum\cite{Sch98}.  SPE's are mostly detected under
conditions of extreme resonance, whereas CDE's and SDE's can be observed at
incident photon energies far above the effective band gap\cite{Sch96}.
This helps disentangle SDE's from SPE's, which are in the same
energy range at small spin magnetizations.

For a QD in the $xy$ plane submitted to a static magnetic field $B$ in the
$z$ direction, SDE's may involve electronic spin-flips or not. The
later excitations are referred to as longitudinal SDE's, and the former as
transverse SDE's. This means that
SDE's are caused by one-electron excitation operators of the kind

\begin{eqnarray}
F_z &=& \sum_{i} f(\vec{r}_i)\, \sigma_z^i \,\,\,\,\,\,{\rm longitudinal}\,\,
 \,\,\,\, (\Delta S_z = 0)
\nonumber
\\
& &
\label{eq1}
\\
F_{\pm} &=& \sum_{i} f(\vec{r}_i)\, \sigma_{\pm}^i \,\,\,\,\,\,{\rm transverse}
\,\, \,\,\,\, (\Delta S_z = \pm 1),
\nonumber
\end{eqnarray}
where $S_z$ is the  $z$ component of the spin of the dot, and $\sigma_z$
is the $z$ component of the Pauli matrix vector.
$\sigma_{\pm} = (\sigma_x \pm \imath \,\sigma_y)/2$ are the spin-flip operators

\begin{equation}
\sigma_+|\uparrow \rangle = \sigma_- | \downarrow \rangle = 0 \,\, ; \,\,
\sigma_+ |\downarrow \rangle =  | \uparrow \rangle \,\, ; \,\,
\sigma_- |\uparrow \rangle =  | \downarrow \rangle .
\label{eq2}
\end{equation}

Due to rotational invariance in spin space,
longitudinal and transverse SDE's are degenerated at $B = 0$ if
$S_z = 0$ in the ground state (gs).
When the magnetic field is not zero, rotational invariance is broken by the
Zeeman term. Then if the gs is almost
paramagnetic, i.e. $S_z \sim 0$ as it occurs at even filling factors $\nu$,
the SDE's are expected to split in a simple way:

\begin{equation}
\omega_{\pm}^{tr} = \omega^{lon} \pm g^* \mu_B B \,\, ,
\label{eq3}
\end{equation}
where the superscript $lon(tr)$ indicates the longitudinal(transverse) character
of the mode, and the subscript $\pm $ corresponds to the two possible spin flip
transitions. $g^*$ is the effective
gyromagnetic factor and $\mu_B$ is the Bohr magneton. A thorough discussion
of longitudinal dipole SDE's in quantum dots within time-dependent
local-spin density theory (TDLSDT) has been presented in Refs.
\onlinecite{Ser97,Ser98}.
If the gs has a large $S_z$, i.e. a large spin magnetization
(ferromagnetic gs), as it happens at odd filling factors $\nu$, Eq. (\ref{eq3})
no longer holds and longitudinal and transverse SDE's display dramatic
differences arising from the spin dependence of the electron-hole (e-h)
vertex corrections.

To study transverse dipole spin modes we have resorted to
current density functional theory (CDFT)\cite{Fer94,Pi98}
together with  its time-dependent version (TDCDFT)\cite{Lip98}.
For the physics we aim to describe here, local-spin density theory
 (LSDT) and CDFT  sensibly yield the same results.
CDFT is expected to be more reliable than LSDT at high magnetic
fields\cite{Vig88}, but the residual interaction in the
longitudinal spin channel is very cumbersome to work out within TDCDFT.
For this reason, whenever we have had to work out the
longitudinal spin response for the sake of comparison with the
transverse one, we have resorted to TDLSDT.
We will see that the spin dipole response is dominated by the transverse
component, especially for filling factors smaller than 2.
In the following we shall use the terms density functional theory (DFT)
and time-dependent density functional theory (TDDFT) when  the
statement  applies to either version of the general method.

The spin and density response of the two dimensional electron gas (2dEG)
has been thoroughly studied by Kallin and Halperin\cite{Kal84} and
MacDonald\cite{Mac85}. As these authors, we have mainly addressed the
response of quantum dots corresponding to integer filling factor gs's.
We have considered only three cases whose ground state is the finite size
analogue of a partially filled 2dEG configuration between $\nu$ = 1 and
2.
One should regard these results as qualitative, since the corresponding
ground states are believed to be very complicated, strongly correlated
ones, and the use of TDDFT to describe them may be questioned.
Nonetheless, these results display the gross trends of the excitation
spectrum, and for that reason we have considered them here.

\section{Results}

We have taken as a case of study a circularly symmetric QD
of radius $R_{dot} \sim$ 164  nm
made of $N = 210$ electrons in a GaAs-AlGaAs heterostructure.
The gs of this dot has been described in detail in
Ref. \onlinecite{Pi98}.
Throughout this work, we shall be using effective atomic units, whose
definition and value for GaAs can be found for example in that
reference.

A major  advantage of
considering a rather large QD for the present discussion
is that several
integer filling factor gs configurations can be identified
as a function of $B$. This  allows to discuss the influence
of the paramagnetic or ferromagnetic character
of the gs on the excitation spectrum.
The large number of relatively close single particle (sp)
levels that one has to handle to obtain the gs and strength functions
poses some technical problems, and is the token to pay for
microscopically study large QD. In this work we have used a small
temperature
$T \leq 0.1$ K. This facilitates the calculations while it does not introduce
any appreciable thermal effect in the results.

In the following we limit the analysis to the most interesting dipole
mode, for which the $f(\vec{r}_i)$ in Eq. (\ref{eq1}) is $x_i$.
Furthermore,
to take advantage of the imposed circular symmetry, we have considered
as dipole
operators the following ones:

\begin{eqnarray}
D_{\pm1, 0} &=& \frac{1}{2}\sum_{j} r_j \,e^{\pm \imath\, \theta_j}
\sigma_z^j
\,\,\, (\Delta S_z = 0)
\nonumber
\\
& &
\label{eq4}
\\
D_{\pm1, \pm} &=& \frac{1}{2}\sum_{j} r_j \,e^{\pm \imath \,\theta_j}
\sigma_{\pm}^j
\,\, \,(\Delta S_z = \pm 1) \,\, ,
\nonumber
\end{eqnarray}
as well as the  combinations:
\begin{eqnarray}
\sum_{j} x^j \sigma_z^j &=& D_{+1, 0} + D_{-1, 0} \,\,\,
(\Delta S_z = 0)
\nonumber
\\
& &
\label{eq5}
\\
\sum_{j} x^j \sigma_{\pm}^j &=& D_{+1, \pm} + D_{-1, \pm} \,\,\,
(\Delta S_z = \pm1)
\,\,\, ,
\nonumber
\end{eqnarray}
where the subscript $\pm 1$ represents the  orbital angular
momentum carried by the excitation, each one corresponding to a
different kind of  left- or right-circularly polarized light.
Within TDDFT, the external probe represented by these operators has also
a harmonic time dependence of frequency $\omega$.

The method we have used to obtain the spin response has been
described in Ref. \onlinecite{Lip98} for the transverse case, and
in Refs. \onlinecite{Ser97,Ser98} for the longitudinal one, so we do
not need to repeat it here. We want to recall
that the mean field entering the Kohn-Sham (KS) equations changes due to
the dynamical spin magnetization induced by the external field.
Within TDDFT, it is assumed that electrons respond as free particles to
the sum of the external plus induced field, which leads to a Dyson-type
integral equation for the correlation function $\chi$

\begin{equation}
 \chi(\vec{r},\vec{r}\,',\omega) = \chi^0(\vec{r},\vec{r}\,',\omega)
+  \int{d\vec{r}_1 d\vec{r}_2\,
\chi^0(\vec{r},\vec{r}_1,\omega)
V_{eh}(\vec{r}_1 , \vec{r}_2)
\chi(\vec{r}_2, \vec{r}\,',\omega) } \; ,
\label{eq100}
\end{equation}
where $\chi^0$ is the free electron correlation function and $V_{eh}$
is the
residual e-h interaction. From $\chi$ it is  possible to obtain
the induced spin magnetization  corresponding to  any of the
above excitation operators
and then determine the dynamical polarizability function
$\alpha(\omega)$. Finally, the strength function $S(\omega)$ is
obtained from the imaginary part of the dynamical polarizability
function as $S(\omega) = \frac{1}{\pi} {\rm Im} [\alpha(\omega)]$.
This procedure has been explicited  in Ref. \onlinecite{Lip98} for the
$D_{+1,-}$ operator.
All these functions can be labeled according to the $\Delta L_z = \pm1$
and $\Delta S_z = 0, \pm 1$ changes induced in the excitation process.
 In the following, we shall analyze the strength
functions corresponding to the operators in Eq. (\ref{eq5}), which
we call $x\sigma_+$, $x\sigma_- $, $x\sigma_z$ and
$x\sigma_x = x (\sigma_+ + \sigma_-)$.

Some characteristics of the strength functions are  easily understood
in terms of the uncorrelated e-h excitations (SPE's) used to  build
the corresponding
$\chi^0(\omega)$, whose basic ingredients are the KS
single particle energies $\epsilon_{nl\sigma}$ and wave functions
$\varphi_{nl\sigma}(\vec{r\,})$ obtained from the solution of the KS
equations which within LSDT read\cite{Ser98}
\begin{eqnarray}
\left[-{1\over 2}\nabla^2\right. &+& {1\over 2}\omega_c\ell_z +
{1\over 8}\omega_c^2r^2 + V^+(r) \nonumber\\
&+& \left. V^H + V^{xc} +(W^{xc}+{1\over 2}g^*\mu_B B)\sigma_z \right]
\varphi_{nl\sigma}(\vec{r}\,)=\epsilon_{nl\sigma}\varphi_{nl\sigma}(\vec{r}\,)
\,\,\, ,
\label{eq101}
\end{eqnarray}
where $V^+(r)$ is the confining potential and
$V^H=\int{d\vec{r}\,\rho(\vec{r}\,')/|\vec{r}-\vec{r}\,'|} $
is the Hartree potential.
$V^{xc}={\partial
{\cal E}_{xc}(\rho, m)/\partial\rho}\vert_{gs}$ and
$W^{xc}={\partial
{\cal E}_{xc}(\rho,m)/\partial m}\vert_{gs}$
are respectively,  the  spin-independent and spin-dependent
 exchange-correlation sp potentials obtained from the
exchange-correlation
energy density ${\cal E}_{xc}(\rho,m)$, and $\rho$ and $m$ are the
electron and spin magnetization densities
$\rho(\vec{r}\,)
\equiv \rho^{\uparrow}(\vec{r}\,) + \rho^{\downarrow}(\vec{r}\,) $,
 $m(\vec{r}\,) \equiv
 \rho^{\uparrow}(\vec{r}\,) - \rho^{\downarrow}(\vec{r}\,) $.
${\cal E}_{xc}(\rho, m)$ has been  constructed
as indicated in Ref. \onlinecite{Ser98}.
Within CDFT, the KS equations are more cumbersome to write down
\cite{Fer94}.
Still, they have a similar structure, and what one has to keep
in mind for the discussions that follow is the existence of an
exchange-correlation potential which plays the role of $V^{xc}$,
 and another exchange-correlation potential  which
plays the role of $W^{xc}$ in the above equation.

One should be  aware that the
residual e-h interaction may change drastically the
picture extracted from the uncorrelated e-h excitations. Yet, as
a useful guide for the discussion, we collect in Fig. \ref{fig1}
the sp energies as a function of the  angular momentum $l$ for
$\nu$ = 8 to 1, corresponding to $B$ values from 1.29 to 10.28 T.
In the bulk of the dot (small $l$ values), these sp energies  bear the
band characteristics of the 2dEG, having similar $\epsilon_{nl\sigma}$
the
sp states that have the same $s_z$ and yield the same Landau level index
$M \equiv n + (|l| - l)/2$, where $n$ is the sp radial quantum number.
For this
dot, the filling factor $\nu$ does represent the number of occupied
Landau bands,
each one labeled as $(M,\uparrow)$ or $(M,\downarrow)$.
In  Fig. \ref{fig1}, upright
full triangles represent $s_z = \,\uparrow$ sp states, and downright,
empty triangles, $s_z = \,\downarrow$ sp states.
 The horizontal lines represent the electron chemical
potential. All the occupied bands are shown, but
to build the valence space of sp states for calculating the
correlation functions
we have usually considered  more empty states than those shown in the
figure.

It is worth to notice the small energy difference between the
$(M,\uparrow)$ and $(M,\downarrow)$ bands
for even $\nu$ values
$\Delta E_{\downarrow\,\uparrow} \sim  |g^* \mu_B B |$,
which has its origin in the Zeeman term, as compared to the large energy
difference between the same bands if $\nu$ is odd, even if
the applied $B$ is relatively small;
compare for instance the $\nu$ = 7 and 6 cases.
That difference in ferromagnetic gs's
mostly comes from  the  spin-dependent exchange-correlation
  potential
$\Delta E_{\downarrow\,\uparrow} \sim 2 |W^{xc}|$,
which is zero or very small in paramagnetic gs's, and sizeable
in ferromagnetic gs's, largely overcoming the Zeeman energy.
Hartree-Fock (HF)\cite{Gud95} and LSDT or CDFT\cite{Fer94}
yield such large gaps, whereas the Hartree approximation does not.
The role of electron-electron interactions in producing these
gaps when $\nu$ is odd has been stressed in Ref. \onlinecite{Son93}.
We are going to see that the effect of the gap is paramount
 on the transverse spin response.

\subsection{Strength functions}

Figure \ref{fig2}  shows the strength function $S(\omega)$
(solid lines) corresponding to $x\sigma_x$.
For the sake  of clarity, we have decomposed $x\sigma_x$
into its  $x\sigma_+$  and $x\sigma_-$ components, which
are drawn in Figs \ref{fig3} and \ref{fig4} respectively.
The associated free responses, i.e. SPE's,  are represented by
dashed lines. The functions are given in effective atomic  units,
and the frequencies in meV.

Let us first comment on the results corresponding to paramagnetic
gs's in which both spin up and down sp states have a tendency
to be equally populated yielding a small
$S_z$ value. As a consequence, the attractive e-h residual interaction
is weak, and SDE's and SPE's lie in the same energy range.
In both  $x\sigma_+$ and $x\sigma_-$ components,
the strength displays a high energy
structure with a frequency always close to the free strength (see also
Fig.
\ref{fig5}), and a low energy
structure. For paramagnetic gs's, the low energy excitation is a
{\it transverse spin edge mode} built from e-h pairs near the Fermi
level. These pairs can be easily identified in Fig. 1(a) and (b),
as they are at
the intersection of the chemical potential with the Landau bands. The
sp band
structure also explains why the edge mode is more fragmented at low
magnetic fields. For example, at $\nu$ = 8 four e-h pairs,
each one involving quite different sp orbital angular momenta from
the other pairs, are contributing to the $x\sigma_-$
strength, whereas only one pair is contributing at $\nu$ = 2. These
pairs are
weakly correlated among them and the result is an edge mode fourfold
fragmented
at $\nu$ = 8, threefold fragmented at $\nu$ = 6, and so on. This nicely
corresponds to the
number of crossings of the Fermi level with the $(M,\uparrow \downarrow)$
bands in Fig. \ref{fig1}.

In the case of $x\sigma_+$, the edge mode is less fragmented  because
some spin-flip e-h transitions having $\Delta S_z = +1$ are Pauli blocked
by our arbitrary election of $B$ in the positive direction of the $z$
axis
(we recall that one has to have $\Delta L_z = \pm 1$, which cannot
always be
 fulfilled simultaneously with the spin  and edge  conditions).
The lacking of the edge state in the $\nu$ = 4 case is
 due to the particular sp structure around the Fermi level at
$B$ = 2.57 T. This accidental fact has no relevance for the general
discussion.

As anticipated, the e-h residual interaction produces a dramatic effect
when the
gs is ferromagnetic.
Even for moderately intense magnetic fields,
the transverse spin mode emerges in these gs's as a very
collective, undamped excitation whose energy we will see in the next
Subsection depends little on the size of the dot provided it is of the
present size or larger. This is somehow the analogous of Kohn's
theorem\cite{Koh61} for charge density excitations in a QD parabolically
confined, and it has the same physical origin, namely, the exact (or
nearly exact in the spin case) cancelation between the bare and
induced e-h interactions.

It can be seen from Fig. \ref{fig3} that when the dot is fully
polarized at $\nu$ = 1, $x\sigma_+$ no longer excites it
because of Pauli blocking. Even at $\nu$ = 3 ($B$ = 3.43 T),
its strength is very small. On the contrary, the excitation produced by
$x\sigma_-$ displayed in Fig. \ref{fig4} is appreciably redshifted
from the free response. The difference between both situations
reveals the strength of vertex corrections arising from
exchange-correlation terms of the electron-electron interaction,
which within TDDFT\cite{Lip98,Ser98} are the only ones
contributing to dress the free e-h vertex in the spin channel.
This effect is
more sizeable at $\nu$ = 1 when the system is fully polarized.

The low energy peak excited by $x\sigma_-$ is taking almost all the
dipole strength, as we will show later.
Low energy SDE's in ferromagnetic gs's caused either by
$x\sigma_+$ or $x\sigma_-$ {\it are not edge, but bulk spin
excitations}. Again, Fig. \ref{fig1} helps understanding this. For
these odd $\nu$ gs's, the Fermi level is between the $(M_{max},\uparrow)$
and $(M_{max},\downarrow)$ Landau bands, the former being occupied and
the later empty. Although finite size effects distort the bands at the
edge
of the dot which is formed by sp states with  high $l$ values,
it is clear that low energy spin-flip transitions involve sp states whose
energy difference is precisely the energy difference
between the $(M_{max},\uparrow \downarrow)$ bands.
These excitations also occur in the bulk (2dEG). The role of the
residual interaction is clearly visible in  Fig. \ref{fig4}  comparing
the free and TDCDFT strength functions at $\nu$=5 and at $\nu$ = 3,
for example.
One sees that the SPE energies have nothing to do with the low SDE energy
Thus, any free e-h model will be of little
help to quantitatively analyze SDE's in partially polarized QD, the
situation worsening the higher the polarization.

We present in Figure \ref{fig5} a more detailed picture of the spin
excitation in the transverse channel at high $B$, showing
the strength function of $x\sigma_x$ in the $2 > \nu \geq 1$ region.
As in previous figures, the dashed lines represent the free
response. In this range of filling factors, which corresponds to
5.14 T $<  B \leq 10.28$ T, we have found\cite{Pi98} that the
$2 S_z$ value steadily increases\cite{note} from zero to 210,  so the
$x\sigma_x$ strength is essentially that of $x\sigma_-$
already discussed.
The interesting new feature in Fig. \ref{fig5} is the
structure of the high frequency
peaks. They are two orders of magnitude less intense than the
low energy ones, which thus exhaust most of the strength.
Of the high energy peaks, the higher ones are excited by
$x\sigma_-$, and the lower ones by
$x\sigma_+$ (obviously, high energy transitions caused by $x\sigma_+$
are blocked only when the system is fully polarized).
It can  be seen that these high energy
peaks are little collective, as SPE's and SDE's are quite
similar, and also that the centroid of the $x\sigma_+$ and $x\sigma_-$
peaks roughly follows the same evolution with $B$ as
the cyclotron frequency $\omega_c = e B/ m c$ does. The value of
$\omega_c$ is indicated in Fig. \ref{fig5} by  vertical arrows.

When both high energy peaks are clearly visible  in the strength,
as for example at $B$ = 7 T, their splitting is a quantitative
measure of the  spin-dependent exchange-correlation  gap $W^{xc}$,
and its measurement may be the  spectroscopic complement to
experimental gap determinations based on the temperature
dependence of the conductivity\cite{Ush90}.
 This is so because $W^{xc}+ g^* \mu_B B/2$ is directly related
to the energy difference of the
$(M,\uparrow)$ and $(M,\downarrow)$ bands around the Fermi level,
see Eq. (\ref{eq101}).
As the sets of ``parallel'' bands $(M+1,\uparrow \downarrow)$
and $(M,\uparrow \downarrow)$ have the same ``width''
as Fig. \ref{fig1}  indicates,
the splitting between  the high energy peaks would be twice the
energy difference between
the $(M_{max},\uparrow)$ and $(M_{max},\downarrow)$  bands. This is
nicely confirmed in the $B$ = 7 and 9 T cases, for which an
explicit  comparison
is possible (see Fig. 5 of Ref. \onlinecite{Pi98}). We recall that this
comparison is meaningful because of the weak effect of the residual
interaction on the high energy peaks.

The longitudinal high energy peak for these configurations
lies\cite{Ser98}  at
$\omega^{lon} \sim \omega_c$ (see also Fig. \ref{fig7}),  and
an expression similar to Eq. (\ref{eq3}) can be written:

\begin{equation}
\omega_{\pm}^{tr} \sim \omega^{lon} \pm 2\, W^{xc} \,\, .
\label{eq102}
\end{equation}
The validity of Eqs. (\ref{eq3}) and (\ref{eq102}) is a consequence
of the weakness of the residual e-h interaction when they hold
(notice that $g^*$ and $W^{xc}$ are negative).

Figure \ref{fig6} shows the longitudinal ($x\sigma_z$) and transverse
($x\sigma_x$) dipole spin strength functions as dotted and solid
lines, respectively. It can be seen that
in ferromagnetic states, the strength is dominated by the
transverse modes. One should also notice that
 for $\nu \geq 2$, i.e. low $B$, apart from some
fine structure the longitudinal and transverse spin responses
have their main  peaks at quite similar energies.

The paramagnetic
configuration at $\nu$ = 2 ($B$ = 5.14 T) shows the interesting
situation in which a zero spin  gs sustains the
simple result anticipated at the introduction: the transverse
modes are just shifted by the Zeeman energy from the longitudinal
ones. Interestingly too, the low energy $x\sigma_+$ mode has almost
collapsed. This suggests the presence of instabilities in this
particular transverse channel, similar to those found in the
longitudinal spin channel in
the $2 \geq \nu > 1$ region\cite{Ser98}.

When the system is fully polarized, the longitudinal spin and
charge density strengths trivially coincide
and of the spin modes, only the transverse one has a
sense\cite{Kal84,Mac85,Ser98}.
The connection between the peak and Zeeman energies
will be discussed in the next Subsection.

The energies of the more intense, high energy  peaks appearing
in the  $x\sigma_+$ and the $x\sigma_-$ strength functions
are  shown in Fig. \ref{fig7} as a
function of $B$. The cyclotron frequency is also represented.
Solid symbols correspond to even filling factor values,
 and open symbols to
odd filling factor values from $\nu$ = 10 ($B$ = 1.03 T) to
$\nu$ = 1 ($B$ = 10.28 T).
Also drawn are the values of the high energy longitudinal spin
peaks\cite{Ser98} (crosses). The triangles and diamonds correspond
to the $B$ = 7, 8 and 9 T peaks in the right panels of Fig. \ref{fig5}.

In agreement with the preceding
discussion, it is seen that for even filling factors the
energies of the transverse SDE
are $\sim \omega_c \pm g^* \mu_B B$, thus close to
$\omega_c$, whereas for odd filling factors they are well
apart from $\omega_c$ by the large spin-dependent  potential
$W^{xc}$. In all cases, these
peaks correspond to bulk modes involving interband e-h
transitions made of sp states each one belonging to a
Landau band with  different index $M$.

Similarly, Fig. \ref{fig8} collects
the energies of the more intense, low energy  peaks appearing
in the  $x\sigma_+$ (top panel)  and of $x\sigma_-$ (bottom panel)
 strength functions.
Solid and open symbols have the same meaning as in Fig. \ref{fig7}.
The Zeeman energy $E_z = - g^* \mu_B B > 0$ is also represented.
 To emphasize the energy staggering, consecutive $\nu$
points have been connected by a thin line. As we have
already discussed, these modes are  spin edge
modes for paramagnetic gs's, and spin bulk modes for ferromagnetic
gs's.

The SDE corresponding to $x\sigma_-$ is the more interesting
one. It is the only transverse spin mode that appears at high
magnetizations, since the one generated by $x\sigma_+$ is Pauli
blocked. For ferromagnetic gs's, this is an undamped excitation since
it is well apart from the SPE's (see Fig. \ref{fig4}).
Notice that the transverse SDE energy approaches the
Zeeman energy  as $B$ increases. At full polarization
($\nu$ = 1), the energy is
close to $E_z$, but not equal to it. We shall come back to
this point in the following.

\subsection{Sum rules}

Further insight onto the strength functions can be obtained from
the evaluation of sum rules, which are their energy moments.
Some of these moments are easy to obtain, model independent
quantities. We are interested here in sum rules for non-hermitian
excitation operators as the $F_{\pm}$ ones defined in Eq. (\ref{eq1}).
These sum rules have been extensively discussed in
Ref. \onlinecite{Lip89}.

We consider the usual Pauli Hamiltonian $H$ describing an $N$ electron
QD submitted to a constant magnetic field in the $z$ direction
(see for example Ref. \onlinecite{Lip97}) and define the following
sum rules (SR):

\begin{eqnarray}
S_0 \equiv S^-_0 - S^+_0 \equiv
\sum_{n}  \,|\langle n| F_- |0\rangle|^2 -
\sum_{n}  \,|\langle n| F_+ |0\rangle|^2 \,\,
\nonumber
\\
& &
\label{eq6}
\\
S_1 \equiv S^-_1 + S^+_1 \equiv
\sum_{n} \omega_{n0} \,|\langle n| F_- |0\rangle|^2 +
\sum_{n} \omega_{n0} \,|\langle n| F_+ |0\rangle|^2 \,\, ,
\nonumber
\end{eqnarray}
where $|0\rangle$ is the gs of the system and $|n\rangle$ is
an excited state with excitation energy $\omega_{n0}$. Using
closure, it is easy to check that

\begin{eqnarray}
S_0 &=& \langle 0|[F_+, F_- ] |0\rangle
\nonumber
\\
& &
\label{eq7}
\\
S_1 &=& \langle 0|[F_+,[H, F_- ] |0\rangle
\nonumber
\end{eqnarray}
An explicit evaluation of $S_0$ and $S_1$ yields

\begin{eqnarray}
S_0 &=& \int d \vec{r}\, |f(\vec{r}\,)|^2 \,m_0(\vec{r}\,)
\nonumber
\\
& &
\label{eq8}
\\
S_1 &=& \frac{1}{2}\int d \vec{r}\, |\nabla f(\vec{r}\,)|^2 \,
\rho_0(\vec{r}\,)
 - g^* \mu_B B \int d \vec{r}\, |f(\vec{r}\,)|^2 \,m_0(\vec{r}\,)
 \,\, ,
\nonumber
\end{eqnarray}
where $\rho_0(\vec{r}\,)$ is the gs density of the dot, and $m_0(\vec{r}
\,)
\equiv \rho_0^{\uparrow}(\vec{r}\,) - \rho_0^{\downarrow}(\vec{r}\,) $
is the gs
local spin magnetization. These sum rules are  relating properties
of the exact spectrum of the Pauli Hamiltonian to properties of its
exact gs.

$S_0$ and $S_1$ are also fulfilled within TDDFT.
 Indeed, it can be  proved using the techniques
discussed in Ref. \onlinecite{Lip89}, that the TDDFT spectrum is
such that one can obtain them with TDDFT accuracy from Eqs. (\ref{eq8})
using the KS gs.
Not all approximation schemes fulfill these sum rules.  Independent
particle spectra  such as those obtained from HF, KS or Hartree
approximations violate $S_1$. This means that  one would
not obtain the second of Eqs. (\ref{eq8}) using in the second of
Eqs. (\ref{eq7}) the corresponding one-body  Hamiltonian.
For example, the KS spectrum we use to build the SPE's and $\chi^0$
correlation function yields

\begin{equation}
\\
S_1^{KS} = \frac{1}{2}\int d \vec{r}\, |\nabla f(\vec{r}\,)|^2 \,
\rho_0(\vec{r}\,)
 - \int d \vec{r} \,\left[ g^* \mu_B B + 2\, W^{xc}(\vec{r}\,)\right]
 |f(\vec{r}\,)|^2 \,m_0(\vec{r}\,)  \,\, .
\label{eq9}
\end{equation}
Within TDDFT, the effect of the e-h induced
interaction is crucial in restoring  $S_1$:
it exactly cancels the $2\, W^{xc}$ contribution in Eq. (\ref{eq9}).
We want to point out in passing
that in the density channel, the induced interaction  is also
responsible for the fulfillment of Kohn's theorem which would be violated
otherwise.

When the dot
has a large gs spin magnetization, the $x\sigma_+$ term in Eq.
(\ref{eq6})
contributes very little to $S_0$ and $S_1$. Thus, one is left with the
$x\sigma_-$ contribution, which is concentrated in a narrow energy
region
(see right panels in Fig. \ref{fig4}). Under these conditions, it is a
fair approximation to identify $\bar{\omega} \equiv S_1/S_0$ with
the mean energy of the peak displayed in the $x\sigma_-$ strength
function. To proceed further, let us consider that the dot is
fully polarized, i.e., $m_0(\vec{r}\,) = \rho_0(\vec{r}\,)$. Taking
$f(\vec{r}\,) = x$, one gets

\begin{eqnarray}
S_0 &=& S_0^- = \frac{N}{2} \langle r^2 \rangle
\nonumber
\\
& &
\label{eq10}
\\
S_1 &=& S_1^- = \frac{N}{2} - \frac{N}{2} g^* \mu_B B \langle r^2
\rangle
\,\, ,
\nonumber
\end{eqnarray}
where $\langle r^2 \rangle$ is the root mean square radius of the dot

\begin{equation}
\langle r^2 \rangle \equiv \frac{1}{N} \int d \vec{r}\, r^2 \,
\rho_0(\vec{r}\,)
\,\, .
\label{eq11}
\end{equation}
Thus

\begin{equation}
\bar{\omega} = \frac{S_1}{S_0} = - g^* \mu_B B + \frac{1}
{\langle r^2 \rangle}
\,\, .
\label{eq12}
\end{equation}
At $\nu$ = 1, taking for $\rho_0(\vec{r}\,)$ that of the maximum
density droplet
(MDD) state\cite{Mac93}, which is a good approximation for large dots,
we have

\begin{equation}
\bar{\omega} = \frac{S_1}{S_0} = - g^* \mu_B B + \frac{\omega_c}{N+1}
\,\, .
\label{eq13}
\end{equation}
Expressions (\ref{eq12}) or (\ref{eq13}) are the SR estimates of the
transverse dipole SDE at large spin magnetization. Using again the MDD
$\langle r^2 \rangle$ value, we get $S_0^- \sim N(N+1) / 2 \omega_c$.
This
shows that for high spin magnetizations, the squared $x\sigma_-$
 matrix elements in the SDE's are a factor
$\sim N$ stronger than the squared $x$  matrix elements in the CDE's,
whose strength is $\sim N / 2 \omega_c$\cite{Lip97}.

For GaAs, $\omega_c/|g^* \mu_B B| =  2/|g^* m^*|
\sim 68$. We thus have  $\bar{\omega} \sim$ 0.35 meV
 for the $N$ = 210 dot at $\nu$ = 1, whereas the peak energy
is $E_{peak} \sim $ 0.27  meV and  $E_z \sim $ 0.26 meV.
In the limit of a very large dot, $\bar{\omega}$ becomes $E_z$. This
is the
analogous of the 2dEG result\cite{Kal84,Mac85} for the spin-wave
dispersion relation at $q$ = 0 (Larmor's theorem).

So far, we have discussed the response to the dipole $L = 1$ operator.
It
is straightforward to consider the response to a general $L$ mode
operator of the kind $f(\vec{r}\,) \sim r^L\, e^{\pm \imath L \theta}$.
These fields are relevant to study  spin and charge density modes with
well defined angular momentum. Moreover, in Raman experiments one may
face a situation in which the in-plane transferred linear momentum $q$
is small enough, so that $q R_{dot} << 1$ and the plane wave operator
$e^{\imath \vec{q}\,\vec{r}}$ involved in the  excitation process can
be expanded into  $ r^L \,e^{\pm \imath L \theta}$ multipoles, each of
them
probing a well defined angular momentum CDE or SDE\cite{Sch98}. When $q$
cannot be considered as small\cite{Loc96}, the TDDFT response may still be
worked out, fixing $q$ and adding the responses to $f(\vec{r}\,) =
J_L(q r) \,e^{\pm \imath L \theta}$, because
$e^{\imath \vec{q}\,\vec{r}} =
\sum_L \imath^L \,J_L(q r)\, e^{\imath L \theta}$. The number of terms
 in the expansion
may be large depending on the $q$ value, but in principle the method is
of direct applicability.

We finish this Subsection discussing
the SR for $f(\vec{r}\,) = e^{\imath \vec{q}\,\vec{r}}$ in the more
interesting case of full polarization. Since
$|f(\vec{r}\,)|^2 = 1$ and $|\nabla f(\vec{r}\,)|^2 = q^2$, we get from
Eqs. (\ref{eq8})

\begin{eqnarray}
S_0 &=& N
\nonumber
\\
& &
\label{eq14}
\\
S_1 &=&  \frac{1}{2} q^2 N -  g^* \mu_B B N
\,\, .
\nonumber
\end{eqnarray}
The same equations hold for the 2dEG
substituting $N$ by the electron density. In either case,

\begin{equation}
\bar{\omega} = - g^* \mu_B B + \frac{1}{2} q^2
\label{eq15}
\end{equation}

We want to emphasize that Eqs. (\ref{eq14}) and (\ref{eq15}) are
exact, model independent and valid for any $q$ value. Only the
applicability of the Pauli Hamiltonian to describe this
 physical situation has been taken for granted.

At first glance, Eq. (\ref{eq15}) is in contradiction with the
spin-wave dispersion relation of Ref. \onlinecite{Kal84,Mac85}, whose
$q$ dependent term has an electron-electron energy dependence instead
of the 1/2 factor. The difference  stems from the
sp valence space, which is different in both calculations. Indeed,
the sum rule result Eq. (\ref{eq15}) takes into account all possible
intraband and interband sp excitations  induced by the operator
$e^{\imath \vec{q}\,\vec{r}}\, \sigma_- $. Thus, $\bar{\omega}$ is an
average of the low and high energy peaks.
In Refs. \onlinecite{Kal84,Mac85}, the valence space was restricted to
the
filled $(0,\uparrow)$ and empty $(0,\downarrow)$ bands to
specifically address the low energy mode. Their result
can be exactly recovered in the SR approach if one  uses
the same valence space and accordingly, the projection of
$e^{\imath \vec{q}\,\vec{r}}\, \sigma_- $
onto the $M$ = 0 space\cite{Gir84}.
This is the so-called single mode approximation
(SMA)\cite{Gir86}, equivalent to the approach of
Ref. \onlinecite{Kal84,Mac85} in the fully polarized case.

It is easy to seize the effect of the $M > 0$ bands on the
low energy collective mode within TDCDFT.
It suffices to compute the response
limiting the sp valence space to the $M = 0$ bands. In this case, only
the low energy peak appears in the $x\sigma_-$ strength function,
and its energy $\sim 0.30$ meV is denoted by a cross in the bottom panel
of
Fig. \ref{fig8}. The inclusion of the high energy bands changes the
energy
of the transverse SDE in $\sim$ 10 \%.
We have also determined that the high energy
peaks in the $x\sigma_-$ strength contribute around 20 \% to $S_1$.

\section{Summary}

In this work we have thoroughly studied the transverse dipole spin
response in quantum dots. Together with our previous
works\cite{Ser97,Lip98,Ser98}, they provide a detailed account of the
applicability of time-dependent density functional theory to the
description of the dipole response of QD in the charge and spin
 channels. Although
microscopic descriptions of similar complexity exist since
some time ago for the charge density modes in medium size
QD\cite{Bro90,Gud91,Gud95}, the spin density modes in QD had
not been previously addressed.
Besides, we have been able to apply the theory to rather large
dots, similar in size to those 	 investigated
in present experiments\cite{Dem90,Str94,Loc96,Sch96,Sch98}

Among SDE's, the transverse ones are especially relevant; in the
longitudinal channel, the residual interaction is fairly weak,
and the SDE's are Landau damped as they are close to
the SPE's (actually,
the same happens with the transverse modes at low $B$).
In the transverse channel, when the dot has a sizeable magnetization
the position of SDE's is shifted away from the SPE's by
exchange-correlation vertex corrections arising from
electron-electron interactions.
As a consequence, a very collective, dispersionless  SDE emerges.
At large spin magnetization, Pauli's principle plays a
prominent role, blocking the $x\sigma_+$ component of the
transverse strength function, which becomes simpler.

The possibility of carrying the calculations in a large dot
displaying several integer quantum Hall gs's
 has permitted us to disclose the
sensitivity of the transverse response to the applied $B$,
which appears as a strong energy oscillation  with
$B$, and a nearly collapsed low energy $x\sigma_+$ mode at
the $\nu$ = 2 paramagnetic gs. The energy oscillations
are also consequence of the different strength of
vertex corrections in ferromagnetic and paramagnetic
gs's.

We have also derived two model independent sum rules that, on
the one hand, can be used as a control for the analysis
of the experimental data, and on the other hand,
have allowed us to relate our results at full magnetization
to previous works on 2dEG.

\acknowledgements

This work has been performed under Grant Nos. PB95-1249 and
PB95-0492 from DGICYT, and 1998SGR00011 from Generalitat de
Catalunya. A.E. and M.B. (Ref. PR1997-0174) acknowledge support
from the DGES (Spain).

\begin{figure}
\caption{(a): Single-particle energies as a function of orbital angular
momentum $l$ for $\nu$ = 8 to 5. The horizontal solid lines represent
the electron chemical potential. The full, upright triangles represent
$(M,\uparrow)$ bands, and the empty, downright triangles represent
$(M,\downarrow)$ bands. (b): Same as (a) for $\nu$ = 4 to 1.}
\label{fig1}
\end{figure}

\begin{figure}
\caption{Strength function corresponding to  $x\sigma_x$
(solid lines). The dashed lines represent the free strength function.
$S(\omega)$ is in effective atomic units divided by $10^5$.}
\label{fig2}
\end{figure}

\begin{figure}
\caption{Strength function corresponding to  $x\sigma_+$
(solid lines). The dashed lines represent the free strength function.
$S(\omega)$ is in effective atomic units divided by $10^5$.}
\label{fig3}
\end{figure}

\begin{figure}
\caption{Strength function corresponding to  $x\sigma_-$
(solid lines). The dashed lines represent the free strength function.
$S(\omega)$ is in effective atomic units divided by $10^5$.}
\label{fig4}
\end{figure}

\begin{figure}
\caption{Same as Fig. 2 for $2 > \nu \geq 1$. The vertical arrows
indicate the value of $\omega_c$.}
\label{fig5}
\end{figure}

\begin{figure}
\caption{Strength function corresponding to  $x\sigma_x$
(solid line) and $x\sigma_z$ (dotted line).
The strengths are in effective atomic units divided by $10^5$
in the transverse case, and by 5$\cdot10^4$ in the longitudinal case
to make it easier to distinguish them.}
\label{fig6}
\end{figure}

\begin{figure}
\caption{Energies of the more intense high energy peaks  excited by
$x\sigma_+$ (squares) and $x\sigma_-$ (circles) as a function of $B$.
 Solid symbols correspond to even filling factor values, and empty
symbols to odd filling factor values
from $\nu$ = 10 ($B$ = 1.03 T) to $\nu$ =  1 ($B$ = 10.28 T).
Also drawn are the energies of the high energy longitudinal spin peaks
(crosses). The triangles and diamonds correspond to
$B$ = 7, 8 and 9 T.}
\label{fig7}
\end{figure}

\begin{figure}
\caption{Bottom panel: Energies of the more intense low energy peaks
excited by $x\sigma_-$ as a function of $B$ for the same configurations
as in Fig. 7. Also shown is the Zeeman energy (dashed line).
The cross at $\nu$ = 1 is the SDE value obtained from TDCDFT when
the sp valence space is limited to  the $M$ = 0 bands.
Top panel: Same as bottom panel for $x\sigma_+$. Some configurations
are absent due to Pauli blocking.}
\label{fig8}
\end{figure}


\begin{references}
\bibitem{note} This is at variance
with what happens for a small QD in a parabolic potential, for which
$2 S_z$ fluctuates in the same $\nu$ region.
\end{references}
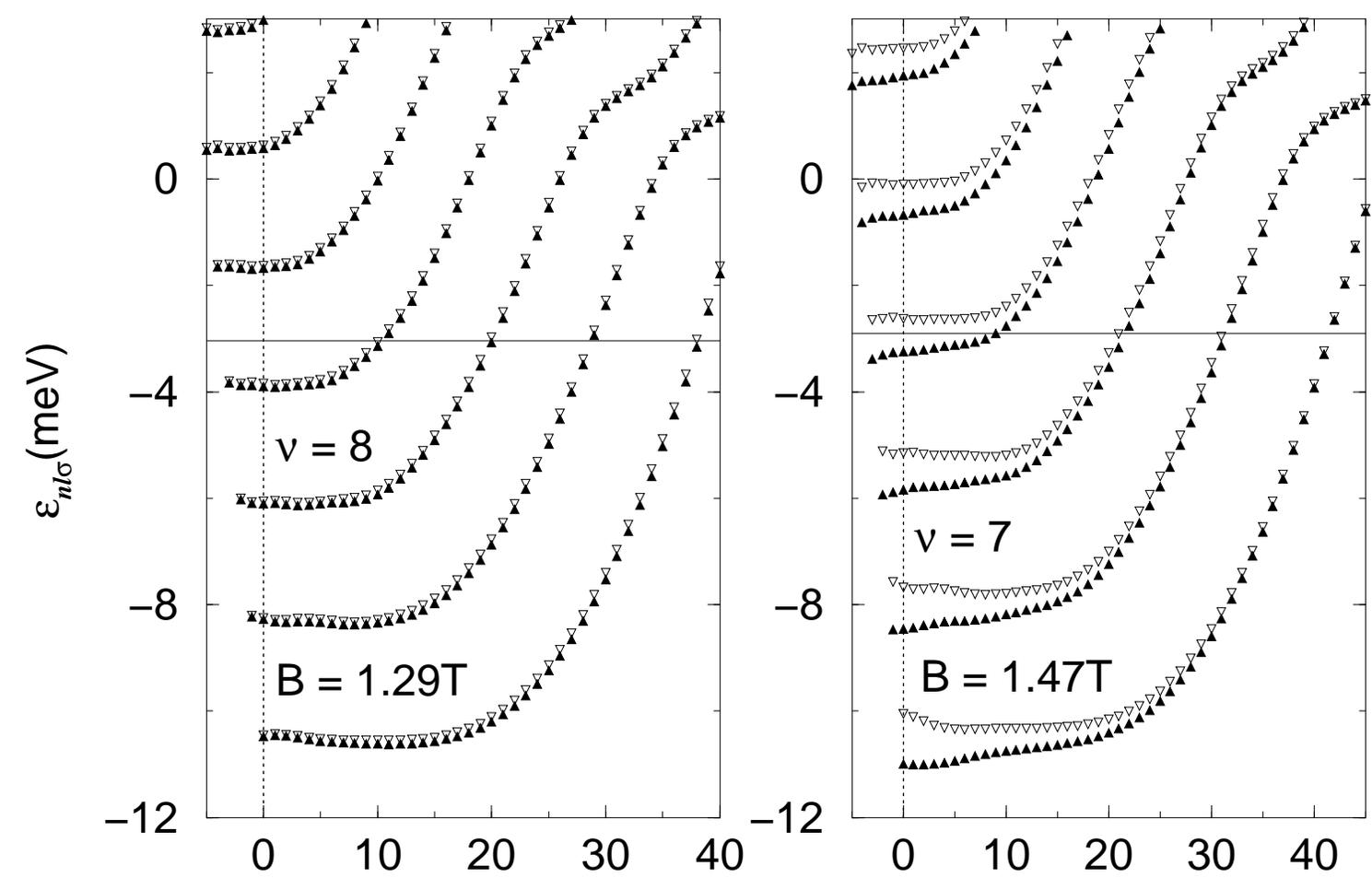
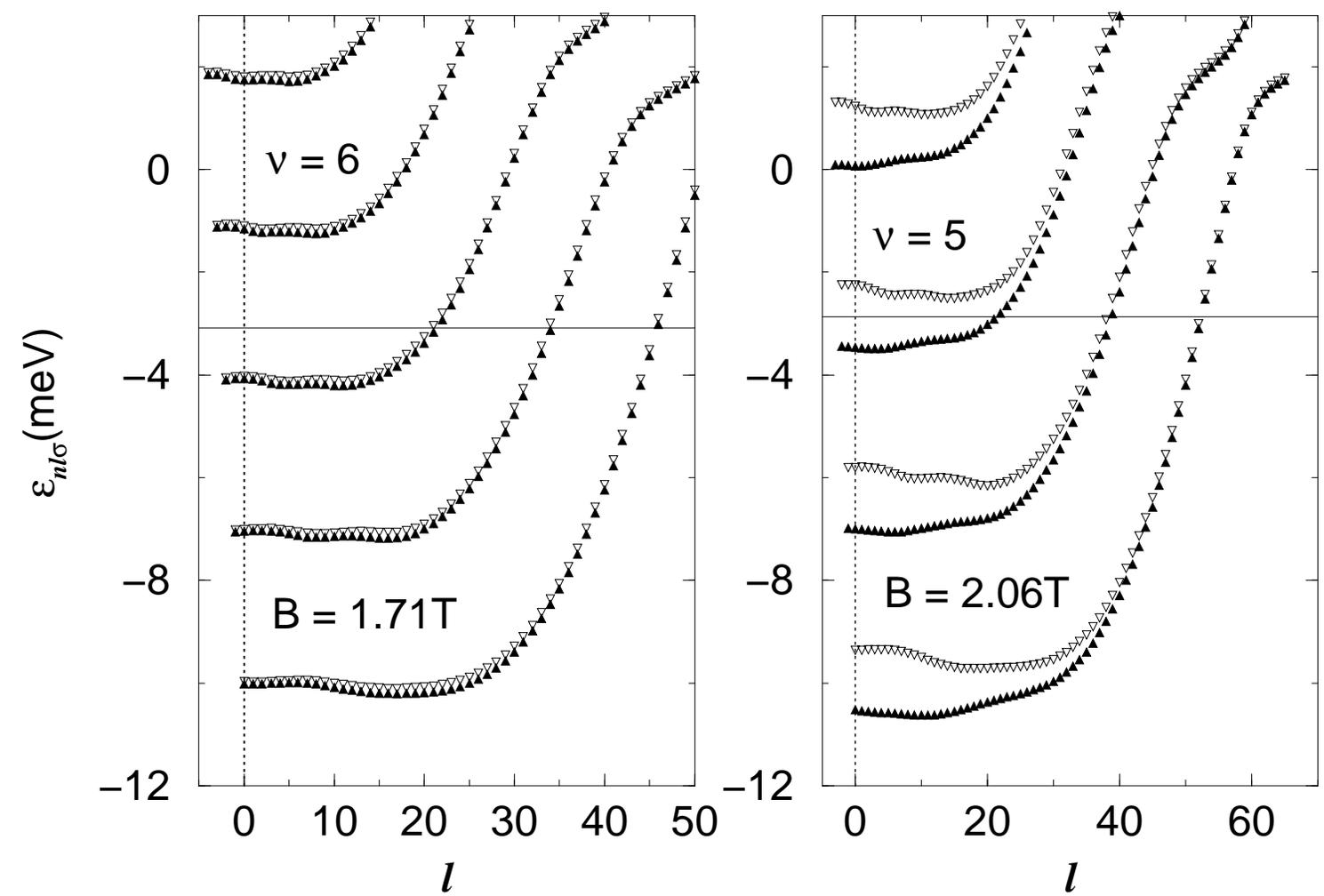
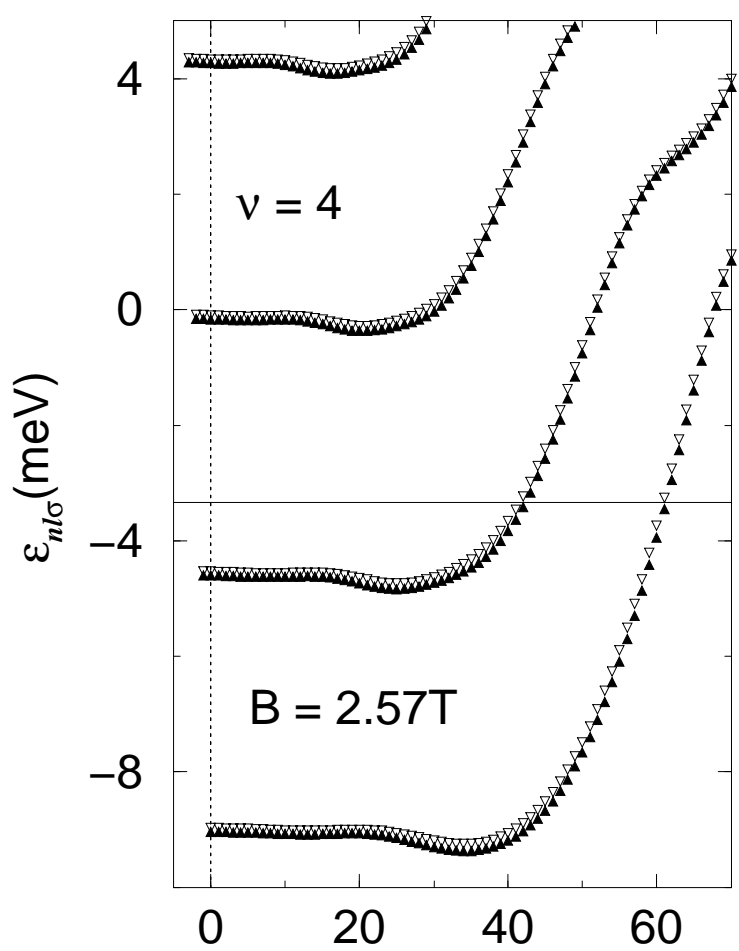
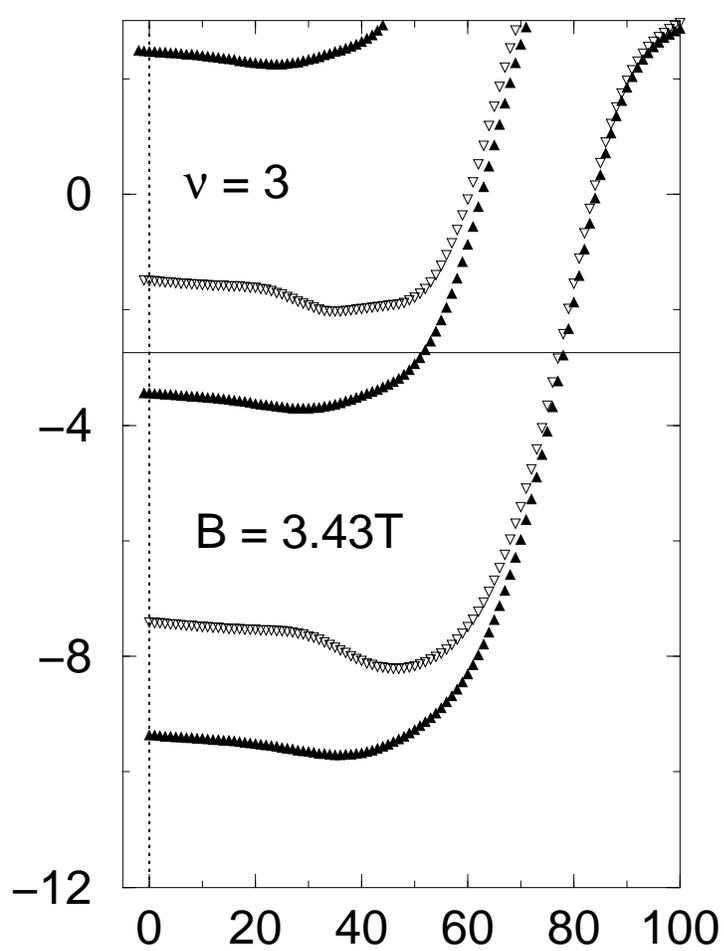
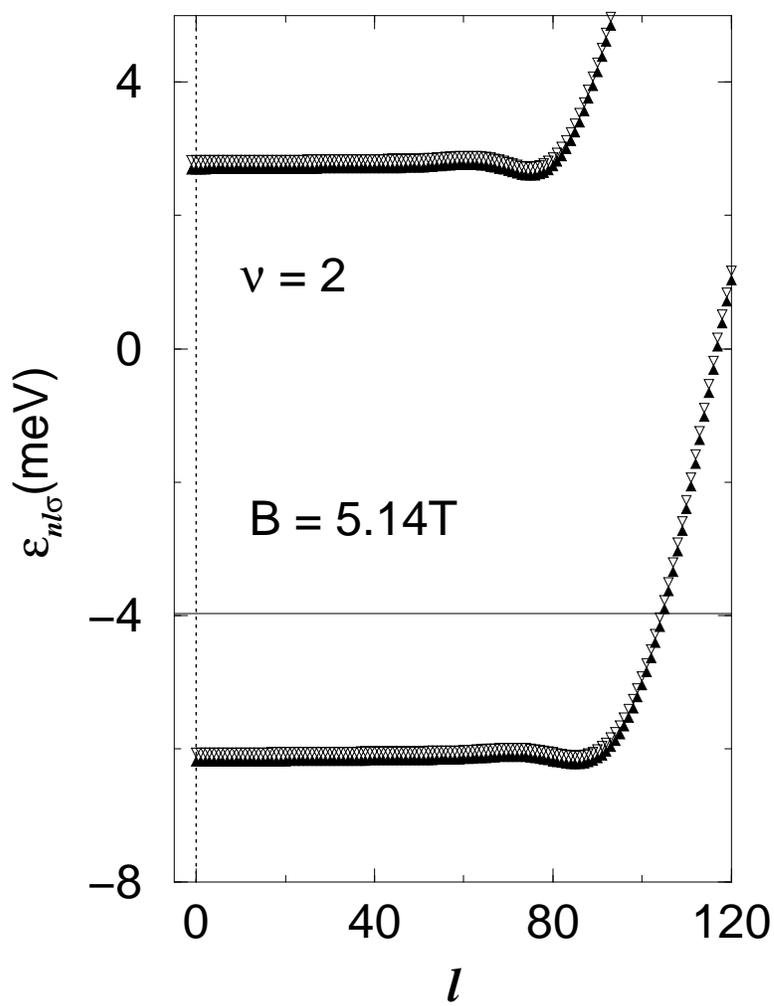
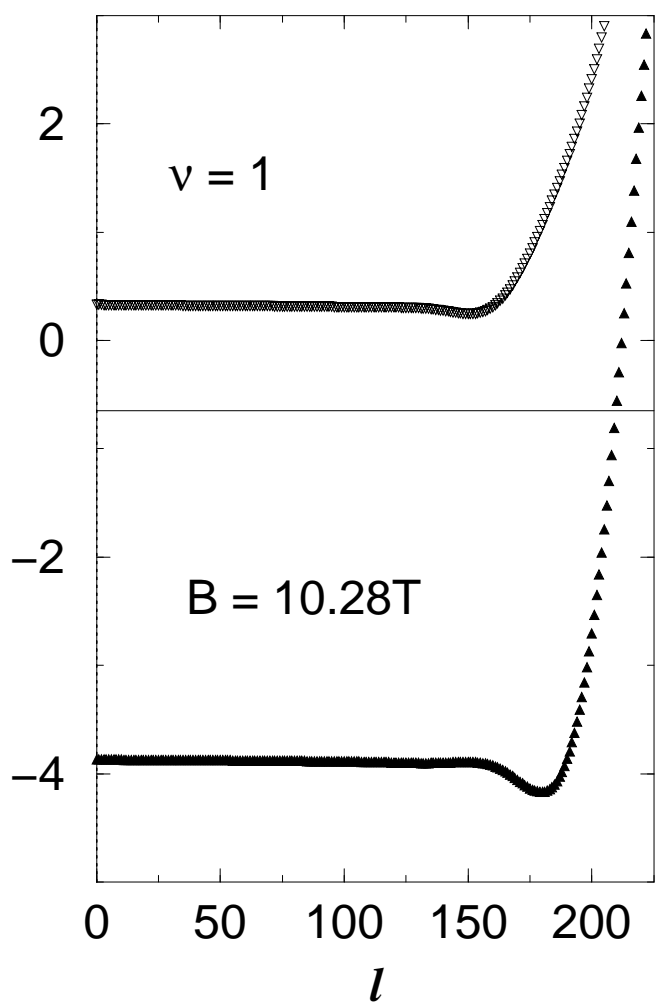
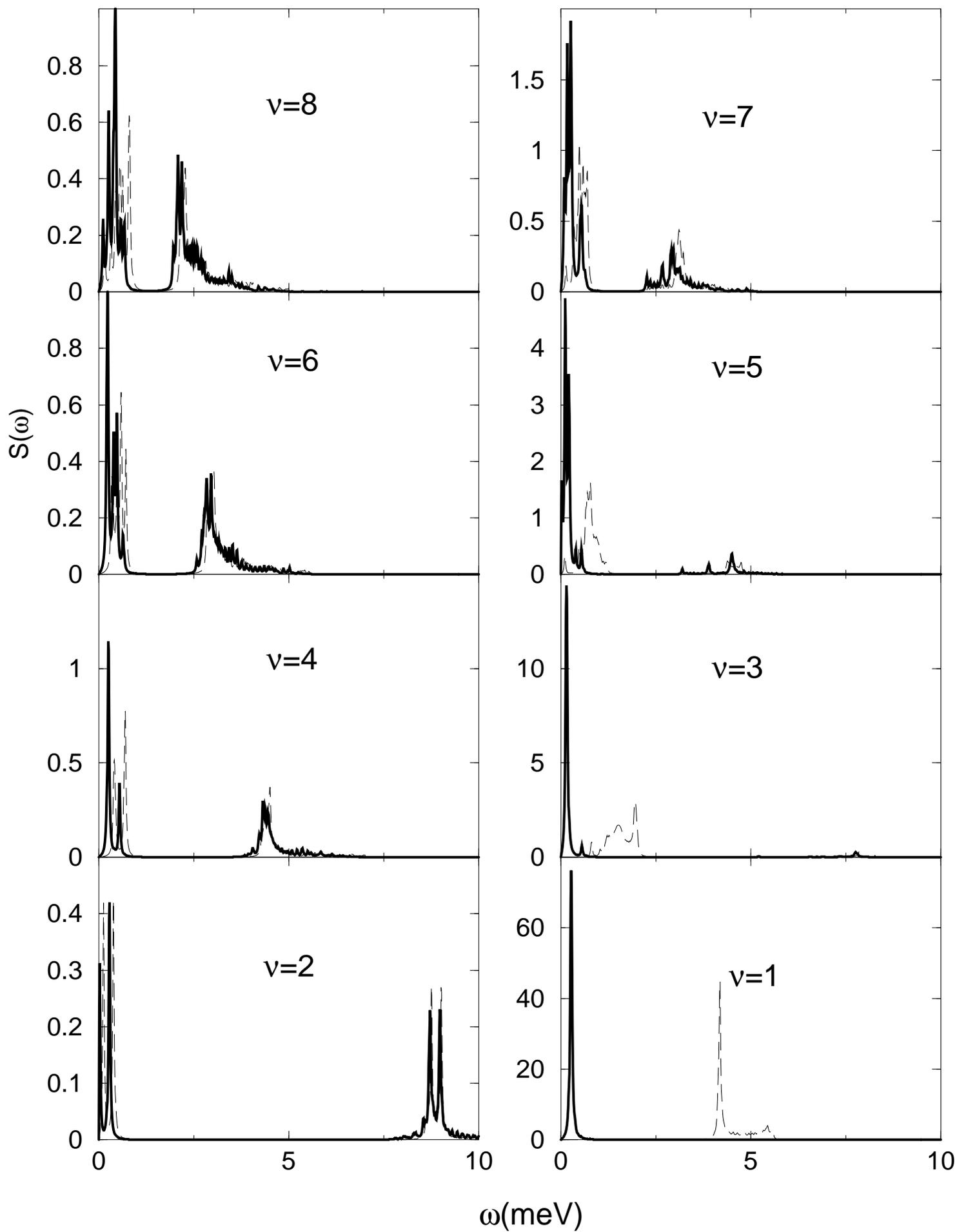
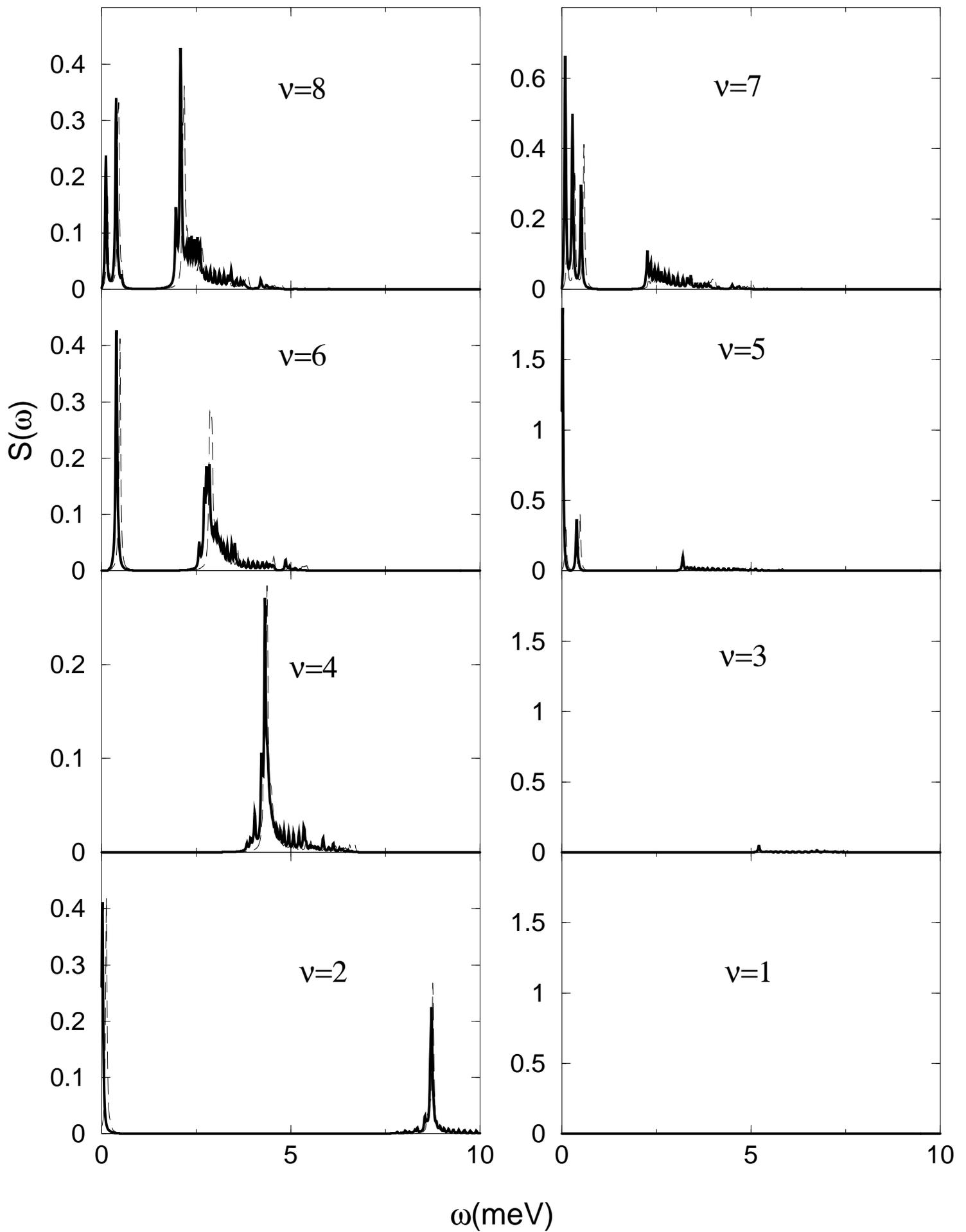
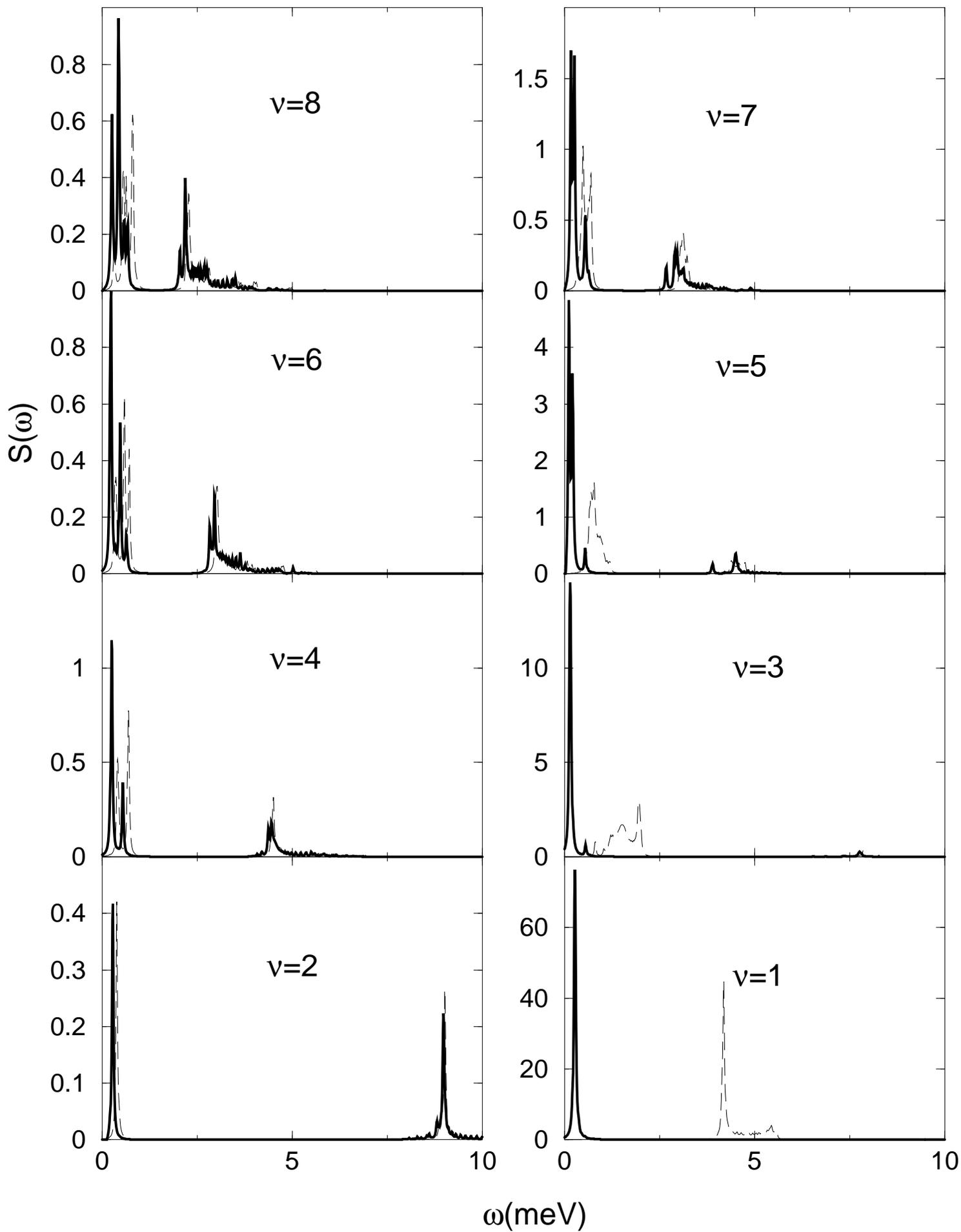
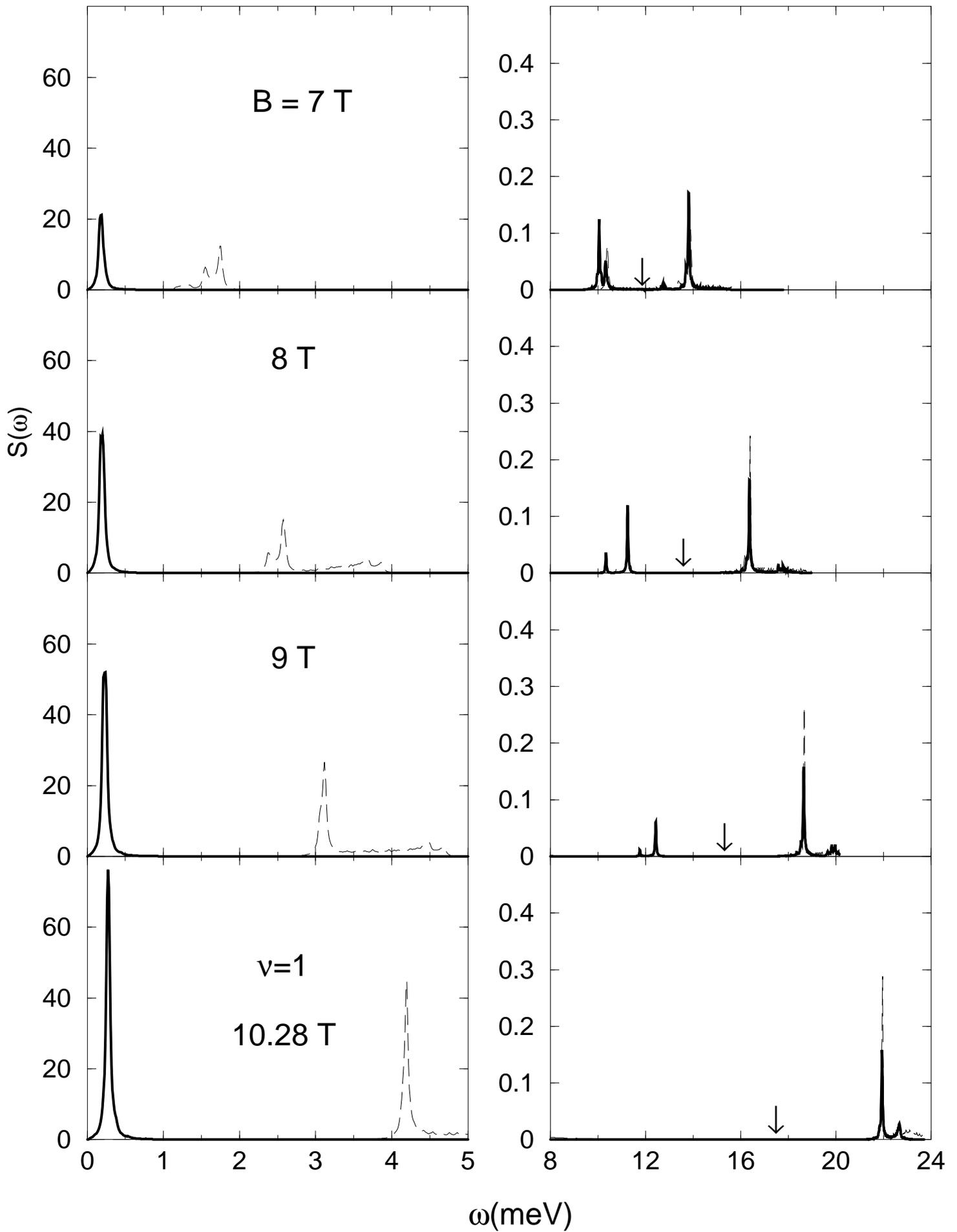
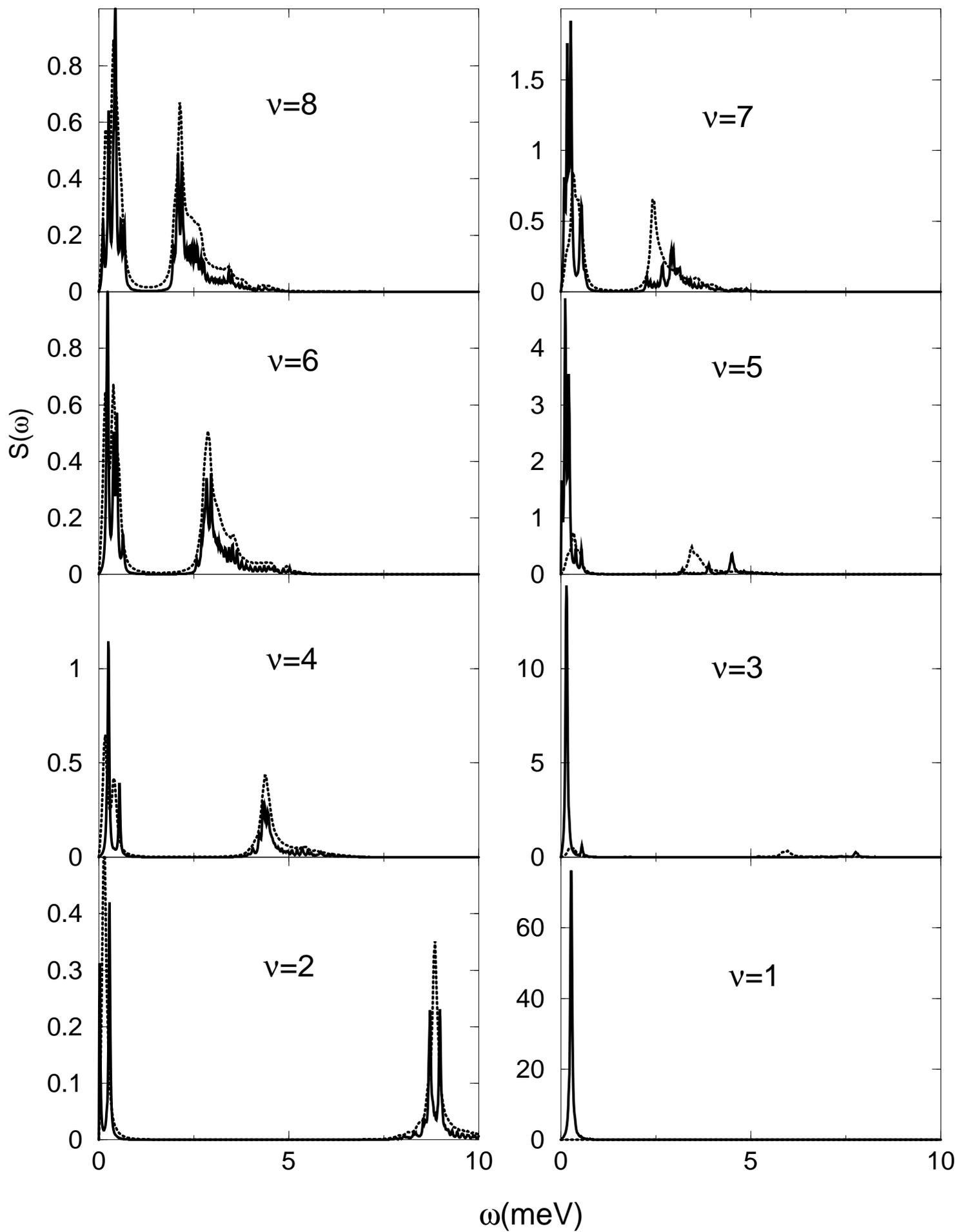
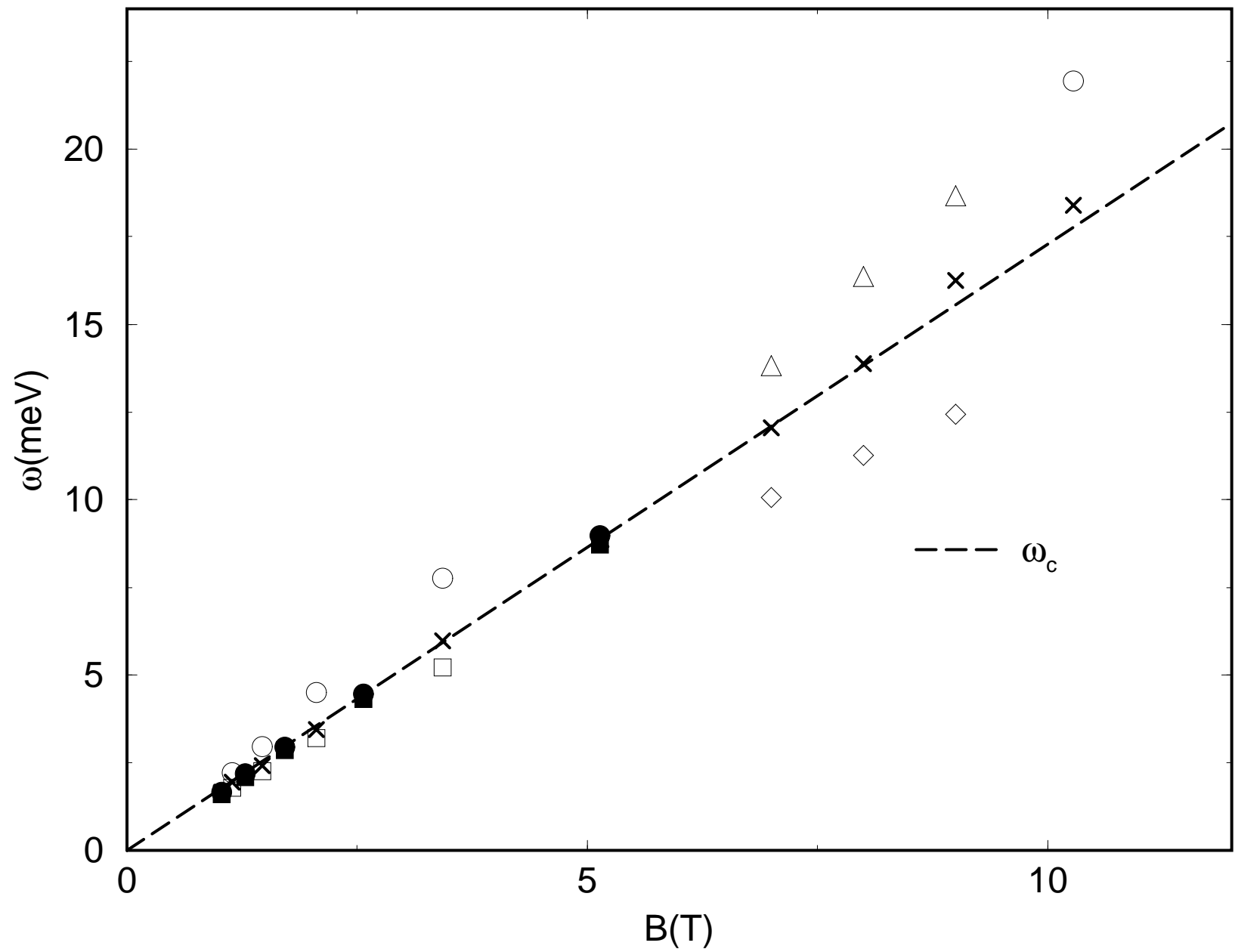
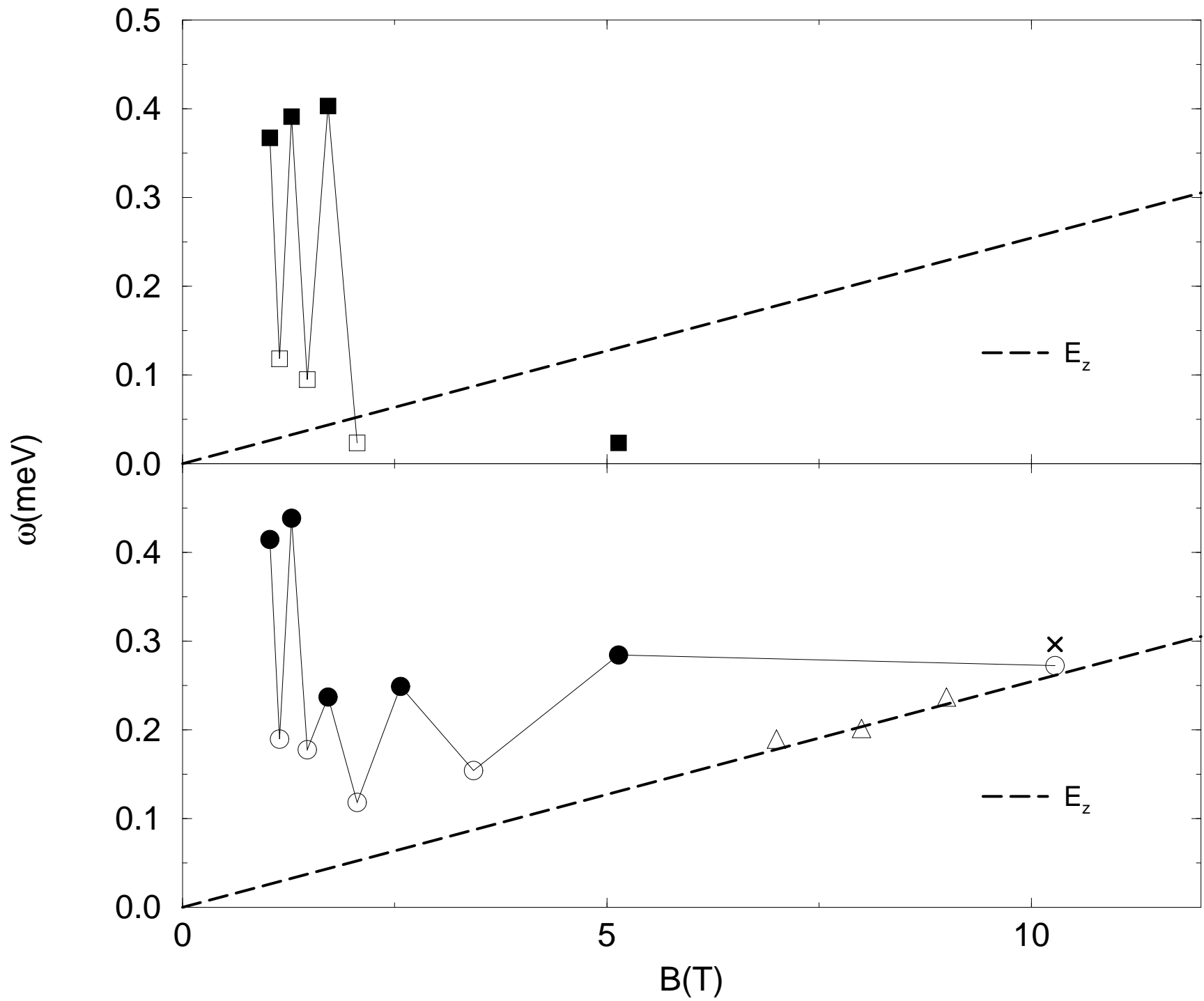
\end{document}